\documentclass[longauth]{aa} 
\usepackage[latin1]{inputenc}
\usepackage[T1]{fontenc}
\usepackage{graphicx}
\usepackage{txfonts}
\usepackage{upgreek}
\usepackage{verbatim}
\usepackage{natbib}
\usepackage{color}

\newcommand{\lsol}{L$_\odot$\,}
\newcommand{\msol}{M$_\odot$\,}
\newcommand{\point}{\,}

\newcommand{\kms}{km$\point${s$^{-1}$}}

\newcommand{\cmcube}{cm$^{-3}$}

\newcommand{\ttp}[1]{$\times 10^{#1}$}

\newcommand{\RA}[3]{$#1^\mathrm{h}#2^\mathrm{m}#3^\mathrm{s}$}
\newcommand{\DEC}[3]{$#1^\mathrm{\circ}#2^\mathrm{\prime}#3^\mathrm{\prime\prime}$}
\newcommand\vlsr{$V_{\rm LSR}$}

\def\hho{H$_2$O}
\def\new#1{\textbf{#1}}
\begin{document}
   \title{Water abundances in high-mass protostellar envelopes: Herschel observations with HIFI\thanks{Herschel is an ESA space observatory with science instruments provided by European-led Principal Investigator consortia and with parti\-cipation of NASA.}}

   \author{
M.G.~Marseille\inst{\ref{inst10}} 
\and F.F.S.~van~der~Tak\inst{\ref{inst10},\ref{inst11}} 
\and F.~Herpin\inst{\ref{inst6}} 
\and F.~Wyrowski\inst{\ref{inst30}} 
\and L.~Chavarr\'{\i}a\inst{\ref{inst6}} 
\and B.~Pietropaoli\inst{\ref{inst6}}
\and A.~Baudry\inst{\ref{inst6}}
\and S.~Bontemps\inst{\ref{inst6}}
\and J.~Cernicharo\inst{\ref{inst16}}
\and T.~Jacq\inst{\ref{inst5}}
\and W.~Frieswijk\inst{\ref{inst10}}
\and R.~Shipman\inst{\ref{inst10}}
\and E.F.~van~Dishoeck\inst{\ref{inst1},\ref{inst2}}
\and R.~Bachiller\inst{\ref{inst12}}
\and M.~Benedettini\inst{\ref{inst13}}
\and A.O.~Benz\inst{\ref{inst3}}
\and E.~Bergin\inst{\ref{inst14}}
\and P.~Bjerkeli\inst{\ref{inst9}}
\and G.A.~Blake\inst{\ref{inst15}}
\and J.~Braine\inst{\ref{inst6}}
\and S.~Bruderer\inst{\ref{inst3}}
\and P.~Caselli\inst{\ref{inst4},\ref{inst5}}
\and E.~Caux\inst{\ref{inst47},\ref{inst48}}
\and C.~Codella\inst{\ref{inst5}}
\and F.~Daniel\inst{\ref{inst16}}
\and P.~Dieleman\inst{\ref{inst10}}
\and A.M.~di~Giorgio\inst{\ref{inst13}}
\and C.~Dominik\inst{\ref{inst17},\ref{inst18}}
\and S.D.~Doty\inst{\ref{inst19}}
\and P.~Encrenaz\inst{\ref{inst20}}
\and M.~Fich\inst{\ref{inst21}}
\and A.~Fuente\inst{\ref{inst22}}
\and T.~Gaier\inst{\ref{inst31}}
\and T.~Giannini\inst{\ref{inst23}}
\and J.R.~Goicoechea\inst{\ref{inst16}}
\and Th.~de~Graauw\inst{\ref{inst10}}
\and F.~Helmich\inst{\ref{inst10}}
\and G.J.~Herczeg\inst{\ref{inst2}}
\and M.R.~Hogerheijde\inst{\ref{inst1}}
\and B.~Jackson\inst{\ref{inst10}}
\and H.~Javadi\inst{\ref{inst31}}
\and W.~Jellema\inst{\ref{inst10}}
\and D.~Johnstone\inst{\ref{inst7},\ref{inst8}}
\and J.K.~J{\o}rgensen\inst{\ref{inst24}}
\and D.~Kester\inst{\ref{inst10}}
\and L.E.~Kristensen\inst{\ref{inst1}}
\and B.~Larsson\inst{\ref{inst25}}
\and W.~Laauwen\inst{\ref{inst10}}
\and D.~Lis\inst{\ref{inst26}}
\and R.~Liseau\inst{\ref{inst9}}
\and W.~Luinge\inst{\ref{inst10}}
\and C.~M$^{\textrm c}$Coey\inst{\ref{inst21},\ref{inst27}}
\and A.~Megej\inst{\ref{inst46}}
\and G.~Melnick\inst{\ref{inst28}}
\and D.~Neufeld\inst{\ref{inst29}}
\and B.~Nisini\inst{\ref{inst23}}
\and M.~Olberg\inst{\ref{inst9}}
\and B.~Parise\inst{\ref{inst30}}
\and J.C.~Pearson\inst{\ref{inst31}}
\and R.~Plume\inst{\ref{inst32}}
\and C.~Risacher\inst{\ref{inst10}}
\and P.~Roelfsema\inst{\ref{inst10}}
\and J.~Santiago-Garc\'{i}a\inst{\ref{inst33}}
\and P.~Saraceno\inst{\ref{inst13}}
\and P.~Siegel\inst{\ref{inst31}}
\and J.~Stutzki\inst{\ref{inst43}}
\and M.~Tafalla\inst{\ref{inst12}}
\and T.A.~van~Kempen\inst{\ref{inst28}}
\and R.~Visser\inst{\ref{inst1}}
\and S.F.~Wampfler\inst{\ref{inst3}}
\and U.A.~Y{\i}ld{\i}z\inst{\ref{inst1}}
          }

   \institute{
Leiden Observatory, Leiden University, PO Box 9513, 2300 RA Leiden, The Netherlands\label{inst1}
\and
Max Planck Institut f\"{u}r Extraterrestrische Physik, Giessenbachstrasse 1, 85748 Garching, Germany\label{inst2}
\and
Institute of Astronomy, ETH Zurich, 8093 Zurich, Switzerland\label{inst3}
\and
School of Physics and Astronomy, University of Leeds, Leeds LS2 9JT, UK\label{inst4}
\and
INAF - Osservatorio Astrofisico di Arcetri, Largo E. Fermi 5, 50125 Firenze, Italy\label{inst5}
\and
Universit\'{e} de Bordeaux, Laboratoire d'Astrophysique de Bordeaux, France; CNRS/INSU, UMR 5804, Floirac, France\label{inst6}
\and
National Research Council Canada, Herzberg Institute of Astrophysics, 5071 West Saanich Road, Victoria, BC V9E 2E7, Canada\label{inst7}
\and
Department of Physics and Astronomy, University of Victoria, Victoria, BC V8P 1A1, Canada\label{inst8}
\and
Department of Radio and Space Science, Chalmers University of Technology, Onsala Space Observatory, 439 92 Onsala, Sweden\label{inst9}
\and
SRON Netherlands Institute for Space Research, PO Box 800, 9700 AV, Groningen, The Netherlands\label{inst10}
\and
Kapteyn Astronomical Institute, University of Groningen, PO Box 800, 9700 AV, Groningen, The Netherlands\label{inst11}
\and
Observatorio Astron\'{o}mico Nacional (IGN), Calle Alfonso XII,3. 28014, Madrid, Spain\label{inst12}
\and
INAF - Istituto di Fisica dello Spazio Interplanetario, Area di Ricerca di Tor Vergata, via Fosso del Cavaliere 100, 00133 Roma, Italy\label{inst13}
\and
Department of Astronomy, The University of Michigan, 500 Church Street, Ann Arbor, MI 48109-1042, USA\label{inst14}
\and
California Institute of Technology, Division of Geological and Planetary Sciences, MS 150-21, Pasadena, CA 91125, USA\label{inst15}
\and
Centro de Astrobiolog\'{\i}a. Departamento de Astrof\'{\i}sica. CSIC-INTA. Carretera de Ajalvir, Km 4, Torrej\'{o}n de Ardoz. 28850, Madrid, Spain.\label{inst16}
\and
Astronomical Institute Anton Pannekoek, University of Amsterdam, Kruislaan 403, 1098 SJ Amsterdam, The Netherlands\label{inst17}
\and
Department of Astrophysics/IMAPP, Radboud University Nijmegen, P.O. Box 9010, 6500 GL Nijmegen, The Netherlands\label{inst18}
\and
Department of Physics and Astronomy, Denison University, Granville, OH, 43023, USA\label{inst19}
\and
LERMA and UMR 8112 du CNRS, Observatoire de Paris, 61 Av. de l'Observatoire, 75014 Paris, France\label{inst20}
\and
University of Waterloo, Department of Physics and Astronomy, Waterloo, Ontario, Canada\label{inst21}
\and
Observatorio Astron\'{o}mico Nacional, Apartado 112, 28803 Alcal\'{a} de Henares, Spain\label{inst22}
\and
INAF - Osservatorio Astronomico di Roma, 00040 Monte Porzio catone, Italy\label{inst23}
\and
Centre for Star and Planet Formation, Natural History Museum of Denmark, University of Copenhagen,
{\O}ster Voldgade 5-7, DK-1350 Copenhagen K., Denmark\label{inst24}
\and
Department of Astronomy, Stockholm University, Albania, 106 91 Stockholm, Sweden\label{inst25}
\and
California Institute of Technology, Cahill Center for Astronomy and Astrophysics, MS 301-17, Pasadena, CA 91125, USA\label{inst26}
\and
the University of Western Ontario, Department of Physics and Astronomy, London, Ontario, N6A 3K7, Canada\label{inst27}
\and
Harvard-Smithsonian Center for Astrophysics, 60 Garden Street, MS 42, Cambridge, MA 02138, USA\label{inst28}M Dishoeck
\and
Department of Physics and Astronomy, Johns Hopkins University, 3400 North Charles Street, Baltimore, MD 21218, USA\label{inst29}
\and
Max-Planck-Institut f\"{u}r Radioastronomie, Auf dem H\"{u}gel 69, 53121 Bonn, Germany\label{inst30}
\and
Jet Propulsion Laboratory, 4800 Oak Grove Drive, MC 302-306, Pasadena, CA 91109  U.S.A.\label{inst31}
\and
Department of Physics and Astronomy, University of Calgary, Calgary, T2N 1N4, AB, Canada\label{inst32}
\and
Instituto de Radioastronom\'{i}a Milim\'{e}trica (IRAM), Avenida Divina Pastora 7, N\'{u}cleo Central, E-18012 Granada, Spain\label{inst33}
\and
Department of Earth and Planetary Sciences, Kobe University, Nada, Kobe 657-8501, Japan\label{inst34}
\and
Universit\'{e} Pierre et Marie Curie, LPMAA UMR CNRS 7092, Case 76, 4 place Jussieu, 75252 Paris Cedex 05, France\label{inst35}
\and
Observatoire de Paris-Meudon, LUTH UMR CNRS 8102, 5 place Jules Janssen, 92195 Meudon Cedex, France\label{inst36}
\and
Department of Physics and Astronomy, San Jose State University, One Washington Square, San Jose, CA 95192, USA\label{inst37}
\and
Laboratoire d'Astrophysique de Grenoble, CNRS/Universit\'{e} Joseph Fourier (UMR5571) BP 53, F-38041 Grenoble cedex 9, France\label{inst38}
\and
European Southern Observatory, Karl-Schwarzschild-Str. 2, 85748 Garching, Germany\label{inst39}
\and
Department of Physics, The University of Tokyo, Hongo, Bunkyo-ku, Tokyo 113-0033, Japan\label{inst40}
\and
Department of Physics and Astronomy, University College London, Gower Street, London WC1E6BT\label{inst41}
\and
Department of Physics, The University of Tokyo, Hongo, Bunkyo-ku, Tokyo 113-0033, Japan\label{inst42}
\and
KOSMA, I. Physik. Institut, Universit\"{a}t zu K\"{o}ln, Z\"{u}lpicher Str. 77, D 50937 K\"{o}ln, Germany\label{inst43}
\and
California Institute of Technology, 1200 E. California Bl., MC 100-22, Pasadena, CA. 91125  USA\label{inst44}
\and
Experimental Physics Dept., National University of Ireland Maynooth, Co. Kildare. Ireland\label{inst45}
\and
Microwave Laboratory, ETH Zurich, 8092 Zurich, Switzerland\label{inst46}
\and
Centre d'Etude Spatiale des Rayonnements, Universit\'e de Toulouse [UPS], 31062 Toulouse Cedex 9, France\label{inst47}
\and 
CNRS/INSU, UMR 5187, 9 avenue du Colonel Roche, 31028 Toulouse Cedex 4, France\label{inst48}
	}

   \date{Received xxx; accepted xxx}

 
  \abstract
{}
{
We derive the \new{dense core structure and the} water abundance in four massive star-forming regions \new{ which may help understand the earliest stages of massive star formation.}
}
{
We present Herschel-HIFI observations of the para-H$_2$O\,$1_{11}-0_{00}$ and $2_{02}-1_{11}$ and the para-H$_2^{18}$O $1_{11}-0_{00}$ transitions. The envelope contribution to the line profiles is separated from contributions by outflows and foreground clouds. The envelope contribution is modeled using Monte-Carlo radiative transfer codes for dust and molecular lines (MC3D and RATRAN), with the water abundance and the turbulent velocity width as free parameters.
}
{
While the outflows are mostly seen in emission in high-J lines, \new{envelopes are seen in absorption in ground-state lines}, which are almost saturated. 
The derived water abundances range from 5\ttp{-10} to 4\ttp{-8} \new{in the outer envelopes}. 
\new{We detect cold clouds surrounding the protostar envelope, thanks to the very high quality of the Herschel-HIFI data and the unique ability of water to probe them}. Several foreground clouds are also detected along the line of sight.
}
{
The low H$_2$O abundances in massive dense cores are in accordance with the expectation that high densities and low temperatures lead to freeze-out of water on dust grains. 
\new{The spread in abundance values is not clearly linked to physical properties of the sources.}
}


   \maketitle
%

\section{Introduction}

\new{Massive stars ($\gtrsim$10\,\msol) play a major role} in the interstellar energy budget and the shaping of the Galactic environment \citep{zinnecker2007}. \new{However,} the formation of such high-mass stars is not well understood due to \new{several reasons: they are rare, have a short evolution time scale, they are born deeply embedded, and are far from the solar system.}

\new{The main sequence lifetime of massive stars} is preceded by an embedded phase which subdivides into several classes of objects: massive pre-stellar cores (mPSC), which are local temperature minima and density maxima within dark clouds \new{\citep{sridharan2005}}; high-mass protostellar objects (HMPO), where a central protostar is surrounded by a massive envelope with a centrally peaked temperature and density distribution \new{\citep{vandertak2000a}}; hot molecular cores (HMC), which have larger masses of warm gas and dust, and high abundances of complex organic molecules which have evaporated off dust grains and/or formed by warm gas-phase chemistry \new{\citep{motte2003}}; and ultracompact \ion{H}{II} regions (UCHII), which show large pockets of ionized gas confined to the star \new{\citep{churchwell1990}}. 
A key question is to what extent these phases represent differences in luminosity and/or age, and if all high-mass stars pass through all these phases. 

The water molecule is thought to be a sensitive tracer of physical conditions in star-forming regions, which acts as a natural filter for warm gas \new{because of its large abundance variations between hot and cold regions \citep{vandertak2006}. Moreover, because the dust continuum is strong at the higher frequencies, water lines connecting with the lowest energy levels can be seen in absorption, thus providing an alternative method to probe different depths in the protostellar environment \citep{poelman2007}.} Measurements of the abundance of water are therefore a step toward understanding the energy budget of star-forming regions, and thus of the star formation process itself.

This paper presents water observations performed with the Heterodyne Instrument for the Far-Infrared (HIFI; \citealt{degraauw}) on-board ESA's Herschel Space Observatory \citep{pilbratt}. We use the p-H$_2$O ground-state line and two lines which constrain the excitation and optical depth (Table~\ref{tab:lines}), \new{all three lying at similar frequencies and observed at similar resolution.}
The sources are four massive star-forming regions (Table~\ref{tab:sources}): the HMCs G31.41+0.31 and G29.96$-$0.02 
and the HMPOs W33A and W43-MM1. We compare our results with those for two other regions: the UCHII region DR21 \citep{dr21} and the HMPO W3~IRS5 (\citealt{chavarria}, this volume).

\new{However, the aim is to discover trends in the water line emission for future extended studies, identifying links in the water abundance between the various evolutionary stages of high-mass star formation and to use water as probe of the gas dynamics around protostars. Given the small number of sources and lines observed, it is premature to look for general trends. 
The large amount of upcoming Herschel-HIFI data will help on this issue.}

\begin{table}[ht]
\caption{List of lines.}
\label{tab:lines}
{\scriptsize
\begin{center}
  \begin{tabular}{lccccccc}
\hline
Molecule & Transition & $\nu$ (GHz) & $E_\mathrm{up}$ (K) & $n_\mathrm{crit}$\tablefootmark{a} (\cmcube) & $\sigma_\mathrm{rms}$ (mK) \\
\hline
\hline
H$_2$O & $1_{11}-0_{00}$ & 1113.343 & 53.4 & 1.7\ttp{8} & 40 \\
H$_2$O & $2_{02}-1_{11}$ & 987.927 & 100.8 & 2.1\ttp{8} & 50 \\
H$_2^{18}$O & $1_{11}-0_{00}$ & 1101.698 & 53.4 & 1.7\ttp{8} & 40 \\
\hline
  \end{tabular}
 \end{center}
}
\tablefoot{
\tablefoottext{a}{Values at 20~K from collision rates of \cite{grosjean2003}}
}
\end{table} 

\begin{table}[ht]
\caption{List of sources.}
\label{tab:sources}
{\scriptsize
 \begin{center}
  \begin{tabular}{lccccccc}
\hline
Name & R.A.	& Dec.	& $L$ 		& d\tablefootmark{a} 	& \vlsr \\
     & J2000	& J2000	& (10$^4$ \lsol)& (kpc)			& (\kms) \\
\hline
\hline
G31.41+0.31 & \RA{18}{47}{34.3} & \DEC{-01}{12}{46.0} & 15 & 7.9 & +98.8 \\
G29.96-0.02 & \RA{18}{46}{03.8} & \DEC{-02}{39}{22.0} & 20 & 7.4 & +98.7 \\
W33A & \RA{18}{14}{39.1} & \DEC{-17}{52}{07.0} & 8.5 & 4.0 & +37.5 \\
W43-MM1 & \RA{18}{47}{47.0} & \DEC{-01}{54}{28.0} & 2.2 & 5.5 & +98.8 \\
\hline
  \end{tabular}
 \end{center}
}
\tablefoot{
\tablefoottext{a}{Values from \cite{hatchell2003}, except W43-MM1 \citep{motte2003} and W33A \citep{vandertak2000a}.}
}
\end{table} 


\section{Observations}

\new{The four regions have been} observed with HIFI  on the 3$^\mathrm{rd}$, 4$^\mathrm{th}$ and 6$^\mathrm{th}$ of March 2010 (see Table~\ref{tab:sources}). Spectra were taken in double sideband mode using receivers 4a (p-H$_2$O at 988~GHz) and 4b (p-H$_2$O and p-H$_2^{18}$O at 1113~GHz and 1102~GHz) with $\nu_\mathrm{LO}$ = 980~GHz and $1108$~GHz respectively. The observations are part of the Priority Science Program (PSP) of the Guaranteed-Time Key Program \textit{Water In Star-forming regions with Herschel} (WISH; Van Dishoeck et al., in prep.).

Data were simultaneously taken with the acousto-optical Wide-Band Spectrometer (WBS) and the correlator-based High-Resolution Spectrometer (HRS), in both horizontal and vertical polarizations. \new{This paper focuses on data from the WBS}, which covers 1140~MHz bandwidth at 1.1~MHz spectral resolution ($\sim$0.3~\kms) \new{\citep{roelfsema2010}}.
System temperatures range between 350~K around 1113~GHz and 450~K around 988~GHz; receiver 4a in V polarization shows particularly high values. 
Integration times (ON+OFF) were 193~s for the 1113~GHz and the 1102~GHz lines, and 206~s for the 988~GHz line, for each source, and the rms noise levels reached are  40--50~mK (Table~\ref{tab:lines}). Observations have been reduced with the Herschel Interactive Processing Environment\footnote{\texttt{http://herschel.esac.esa.int/}} (HIPE) version 2.8. The intensity scale is converted to $T_\mathrm{mb}$ using main beam efficiencies of $0.74$.  
The double-side band continuum level has been divided by 2 to make its brightness directly comparable to that of the lines, which are measured in single sideband.


\section{Results}

\begin{figure*}[ht]
 \centering
 \resizebox{0.24\hsize}{!}{\includegraphics{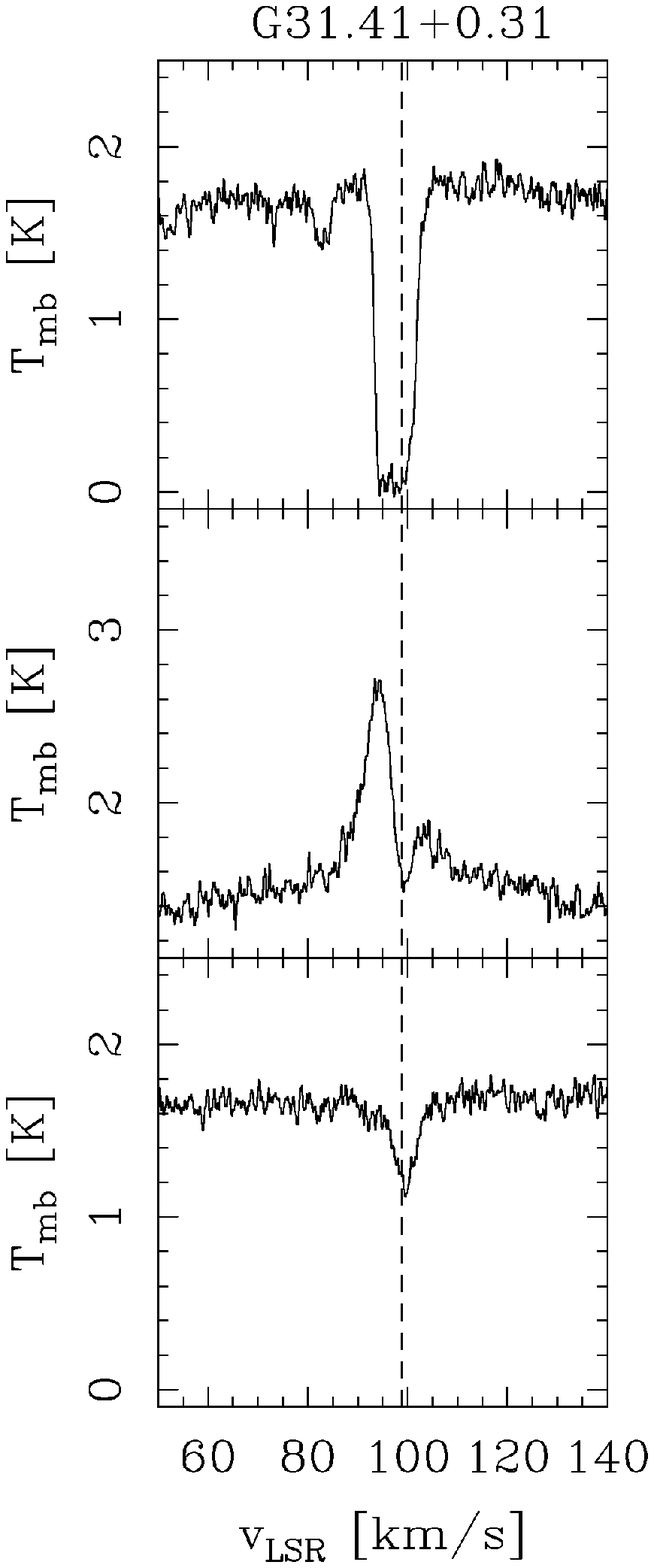}}
 \resizebox{0.24\hsize}{!}{\includegraphics{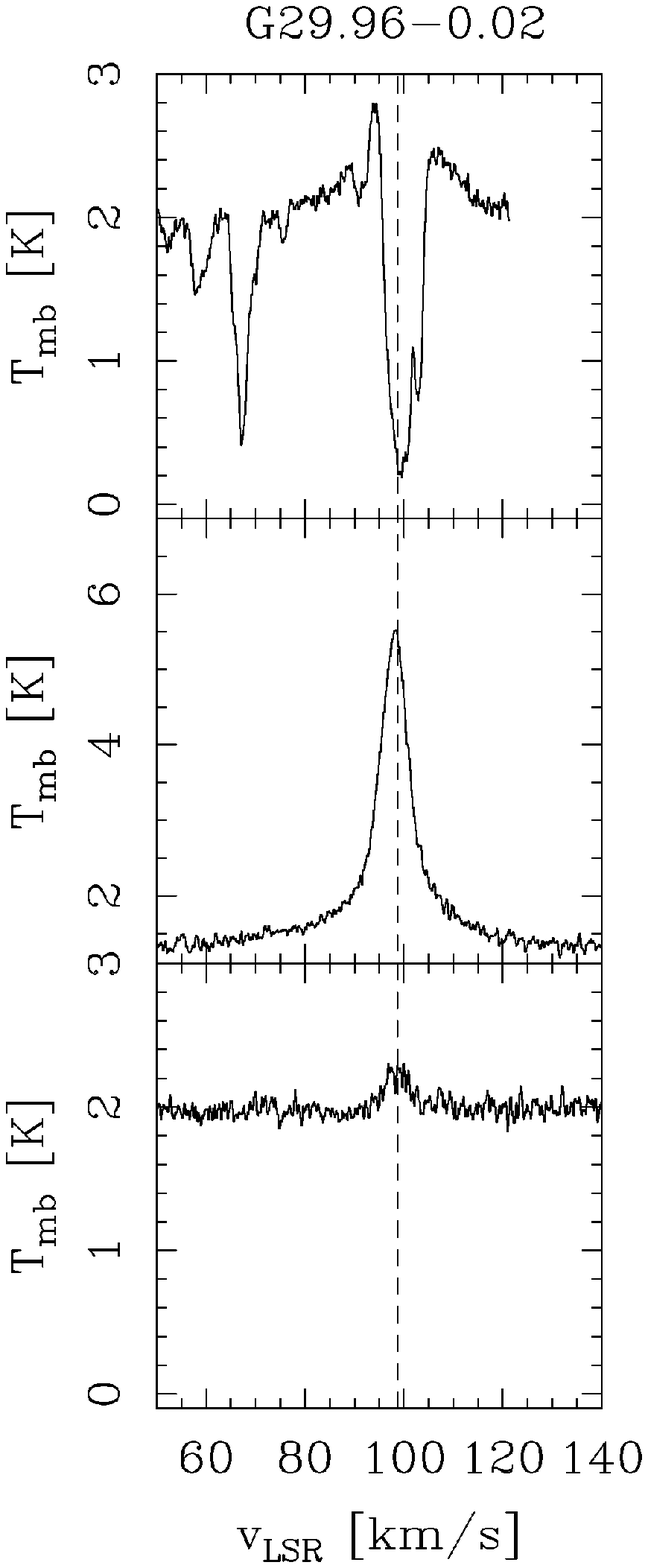}}
 \resizebox{0.235\hsize}{!}{\includegraphics{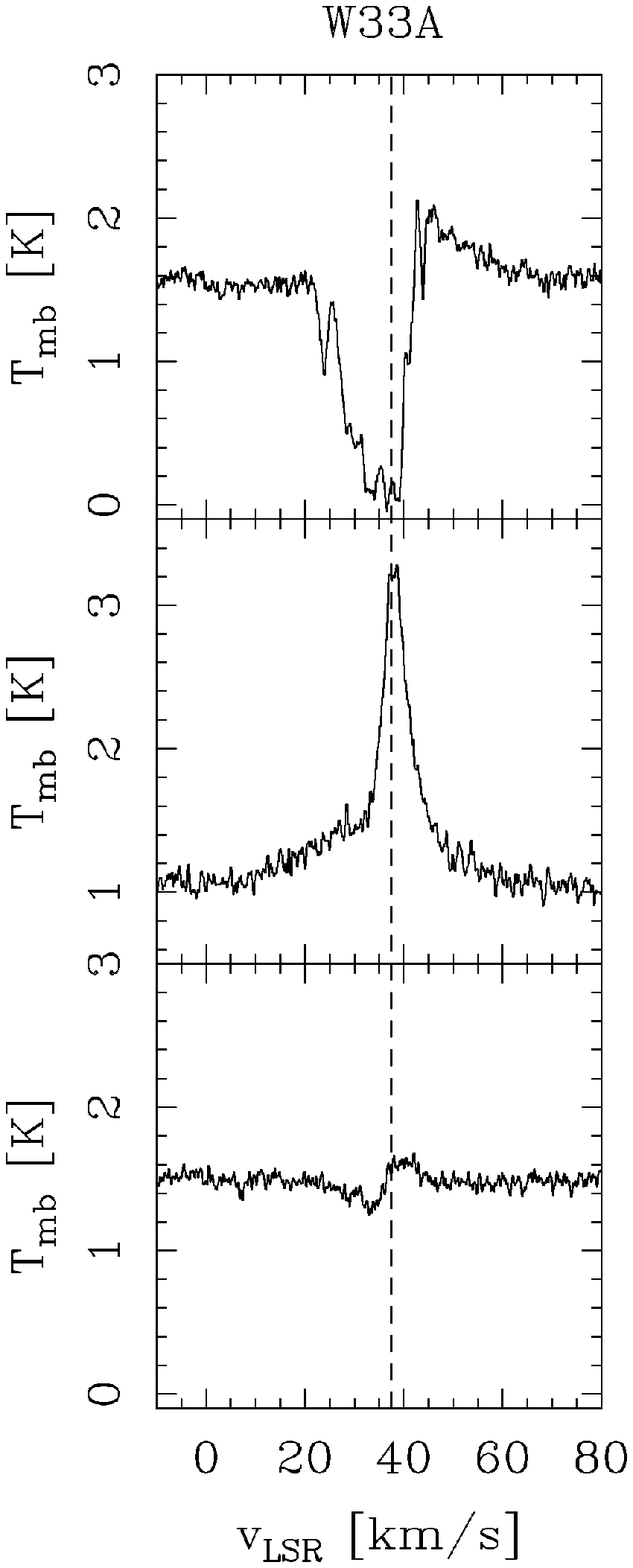}}
 \resizebox{0.24\hsize}{!}{\includegraphics{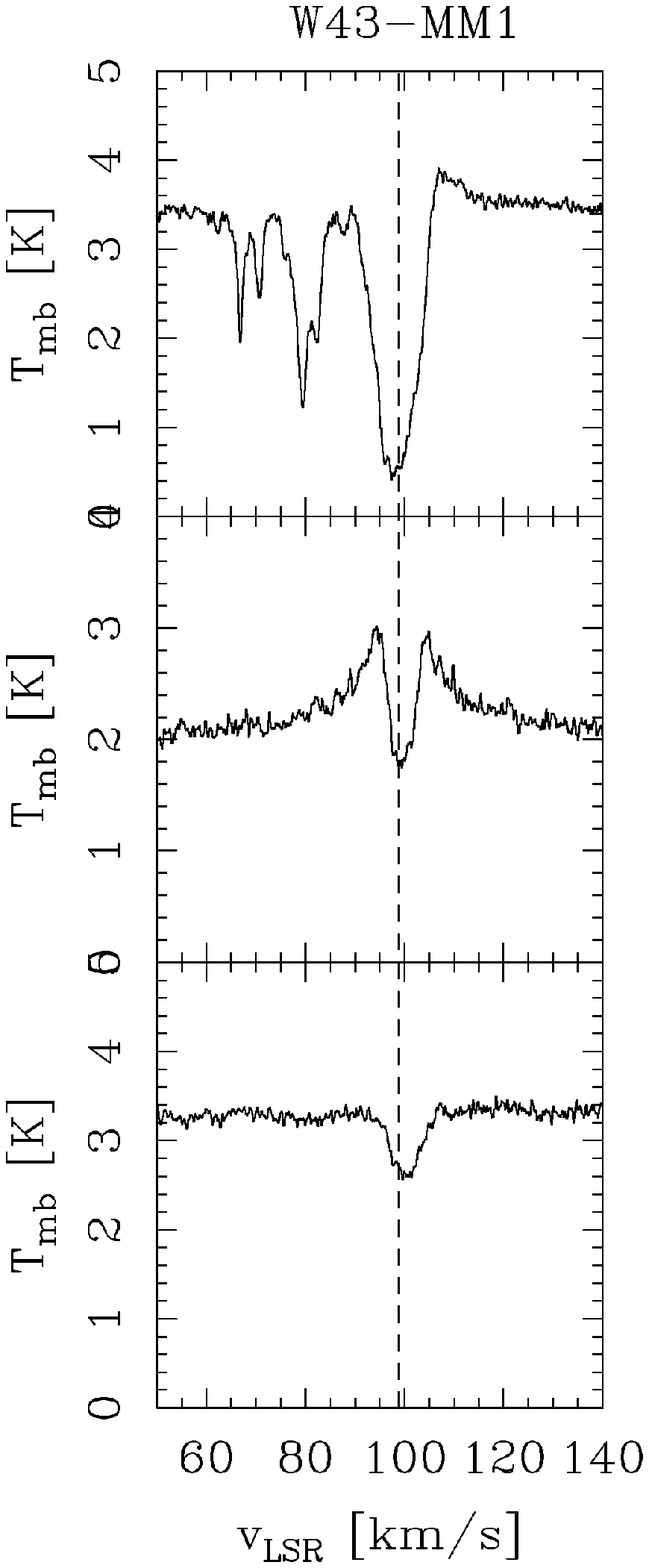}}
 \caption{Herschel/HIFI spectra of the H$_2$O $1_{11}-0_{00}$ (top), H$_2$O $2_{02}-1_{11}$ (middle) and H$_2^{18}$O $1_{11}-0_{00}$ (bottom) lines. Dashed lines drawn at \vlsr.}
 \label{fig:obs}
\end{figure*}

\begin{table*}
 \centering
 \caption{Gaussian decomposition of the line profiles at velocities close to \vlsr.}
 \label{tab:components}
 \begin{center}
  \begin{tabular}{l|ccc|ccc|ccc}
  \hline
Source & \multicolumn{3}{c|}{para-H$_2$O\,($1_{11}-0_{00}$)} & \multicolumn{3}{c|}{para-H$_2$O\,($2_{02}-1_{11}$)} & \multicolumn{3}{c}{para-H$_2^{18}$O\,($1_{11}-0_{00}$)}\\
  \cline{2-4}\cline{5-7}\cline{8-10} 
 & \vlsr\ & $T_\mathrm{mb}$ & $\Delta V$ & \vlsr\ & $T_\mathrm{mb}$ & $\Delta V$ & \vlsr\ & $T_\mathrm{mb}$ & $\Delta V$ \\
  & (\kms) & (K) & (\kms) & (\kms) & (K) & (\kms) & (\kms) & (mK) & (\kms) \\
  \hline
  \hline
G31.41+0.31 	& 95.1 			& 0.94\tablefootmark{*}		& 3.7			& 94.6			& 1.37				& 6.4			& 99.5 	& 0.27\tablefootmark{*} 	& 5.2 \\

		&			&				&			& 99.3			&
0.42\tablefootmark{*}		& 14.0			& 	& 	& \\

&			&				&			& 103.7			&
0.2		& $\sim$40		& 	& 	& \\
  \hline
G29.96-0.02 	& 91.3			& 0.26\tablefootmark{*}				& 3.9			& 97.8			& 
1.10		& 21	 		& 98.5 	& 290 	& 6.0 \\
		& 98.5			& 0.90				& 18.8			& 98.2			&
3.21		& 8.0			& 	& 	& \\
		& 99.4			& 0.99\tablefootmark{*}				& 8.4			& 			&
		& 			& 	& 	& \\
		& 103.2			& 0.40\tablefootmark{*}				& 2.3			& 			&
		& 			& 	& 	& \\
  \hline
W33A 		& 35.9	 		& 0.85\tablefootmark{*}				& 11.0			& 35.5			& 
0.34		& 28			& 34.2	& 0.23\tablefootmark{*}	& 11.8 \\
		& 43.0			& 0.53				& 20.0			& 38.3			&
1.87		& 7.0			& 37.3	& 280	& 8.4 \\
  \hline
W43-MM1 	& 98.7	 		& 0.87\tablefootmark{*}				& 13.5			& 99.6			& 	0.89		& 22			& 99.4 & 0.19\tablefootmark{*} & 8.7 \\
		& 103.3			& 0.43				& 14.0			& 99.7			&
0.31\tablefootmark{*}		& 6.8			&  	&   	&  \\
\hline
  \end{tabular}
 \end{center}
\tablefoot{
\tablefoottext{*}{Absorption lines are indicated in $T_\mathrm{abs}/T_\mathrm{continuum}$ scales.}
}
\end{table*}

\begin{figure*}[ht]
 \centering
 \resizebox{0.9\hsize}{!}{\includegraphics{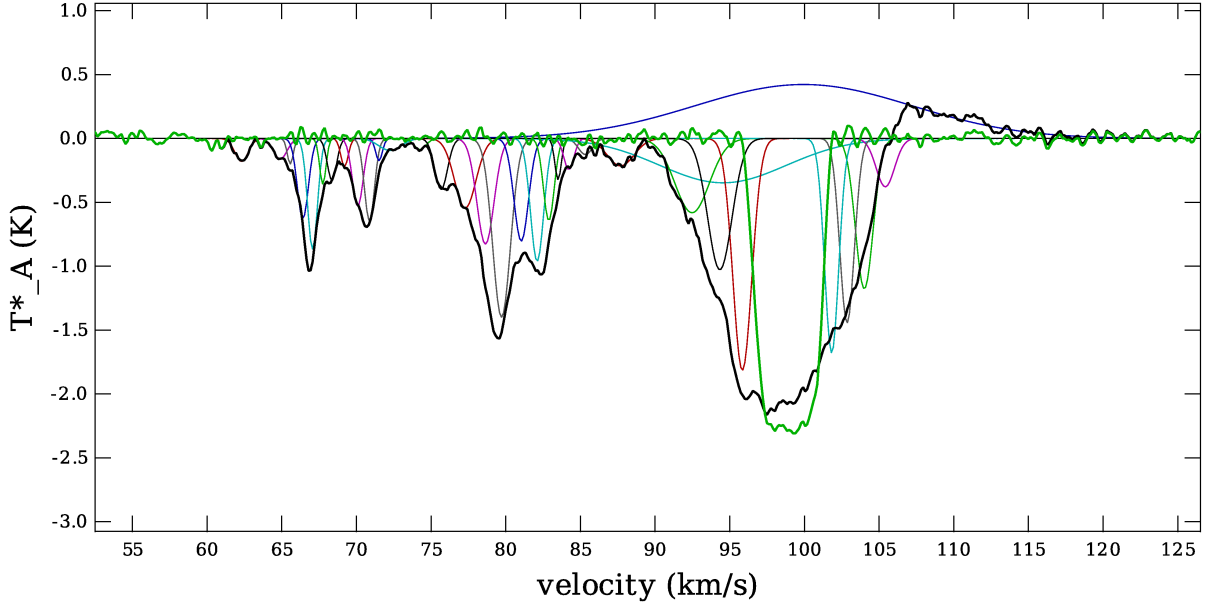}}
 \caption{Extraction of the saturated absorption of para-H$_2$O\,$1_{11}-0_{00}$ line in W43-MM1. Original profile appears in black bold, residual in green bold.} 
 \label{fig:extract-w43-mm1}
\end{figure*}

\new{Observed water lines for the four studied regions are shown in Fig.~\ref{fig:obs}}. The H$_2$O $1_{11}-0_{00}$ line shows an absorption at the systemic velocity (\vlsr) in all sources. In all cases except G31.41+0.31, outflow wings are detected close to the main absorption, with a maximal shift of 3~\kms. These wings are seen in emission, which indicates an origin in hot, low-density \new{($10^3$~\cmcube)} gas \citep{poelman2007}. \new{Absorption features are seen over a large velocity range  in G29.96-0.02, W33A, W43-MM1 and more weakly in G31.41+0.31}. The absorptions at velocity offsets $>$4~\kms\ likely originate in cold foreground clouds on the line of sight to the source. In contrast, the absorption features at smaller velocity offsets \new{are plausibly related} to cold clouds surrounding the dense cores (which other studies call the protostellar envelopes), which are all part of large-scale molecular clouds (see Fig.~\ref{fig:extract-w43-mm1}). \new{Table~\ref{tab:components} presents a Gaussian decomposition of the line profiles around \vlsr; Appendix~A shows Gaussian iterative decompositions of the absorption profiles  of the ground-state transition over the full velocity range, showing a large number of velocity components thanks to the high velocity resolution of the HIFI instrument.}
The absorptions at \vlsr\ are saturated for G31.41+0.31 and W33A and nearly saturated for the other sources, which indicates abundances around $\sim$10$^{-9}$ for the outer cold parts of the massive dense cores \citep{poelman2007}.

The H$_2$O $2_{02}-1_{11}$ line always appears in emission and shows a broad and a narrower velocity component (Fig.~\ref{fig:obs}). In addition, the spectra of G31.41+0.31 and W43-MM1 show two well defined self-absorption features which appear at the source velocity. With its high $E_\mathrm{up}$, this transition mainly traces warm gas, and the presence of these absorption features in G31.41+0.31 and W43-MM1 suggests a higher water abundance in these sources than in G29.96--0.02 and W33A. The components seen in emission have Gaussian shapes, one being wider (FWHM=20--40~\kms) than the other (FWHM=6.4--8.0~\kms). We associate the broad component with high-velocity outflows \new{associated with} the protostar \new{also seen in $1_{11}-0_{00}$ line emission}. This component is symmetric with respect to the source velocity in G29.96--0.02 and W43-MM1,  blueshifted by $2.8$~\kms\ in W33A, and redshifted by $4.4$~\kms\ in G31.41+0.31  (Table~\ref{tab:components}). 
The narrower (hereafter 'medium') component \new{is potentially} associated with shocked surrounding material where water is released in the gas phase. Indeed, shocks occur at the interface between jets and the surrounding dense envelope, with a velocity close to that of the massive dense core. Similar results are found in \cite{kristensen}, \cite{johnstone} and \cite{chavarria}.

The H$_2^{18}$O $1_{11}-0_{00}$ transition is seen in absorption at the source velocity in G31.41 and W43-MM1, which is not saturated. This feature originates from the massive envelope. In G29.96, the pure emission profile of this transition implies a warm diffuse gas origin. Since G29.96 is also the brightest source in the narrower component of the 988~GHz line, we suggest that the H$_2^{18}$O emission is dominated by shocks at the interface between jets and the envelope.
The P-Cygni-like profile for W33A is a mix of both behaviors: the sum of an absorption feature due to the massive core and emission from an outflow. The fact that the emission is seen only on the red-shifted part of the profile is consistent with the outflow components seen in 1113~GHz and 988~GHz lines, which are also more powerful on the red part of the spectrum. 



\section{Discussion and conclusions}

To derive the water abundance in the four massive dense cores we have removed features related to outflows and foreground clouds from the spectrum before any line modeling. 
The high spectral resolution of HIFI is essential in this process, in particular for the absorbers with velocities close to that of the central source.
Studying the H$_2^{18}$O $1_{11}-0_{00}$ transition prior to the others also facilitates to disentangle the \new{envelope} contribution, \new{since this line is not saturated having a lower optical depth than the main H$_2$O isotope.} 


Once the main contribution is extracted, we model its profile according to the method described in \cite{marseille2008}: first, the dust emission from the massive dense core is reproduced with the MC3D radiative transfer code \citep{wolf1999}, including total luminosity and density profile from the literature (power-law index $p = -1.5$); second, the temperature profile obtained is used to model the line emission with the RATRAN code \citep{hogerheijde2000}. The free parameters are: $X_\mathrm{H_2O}$ the molecular abundance relative to H$_2$, and $\varv_\mathrm{turb}$, the turbulent velocity width. 

Good fits are obtained for the H$_2^{18}$O $1_{11}-0_{00}$ transition, which is not saturated unlike the H$_2$O lines. The fitting considers both the line strength (area and width) and the profile shapes. We have computed a grid of  $X_\mathrm{H_2O}$ and $\varv_\mathrm{turb}$ values, adapting step by step the grid around the best $\chi^2$.  Using a $\mathrm{^{16}O/^{18}O}$ ratio of 500, we proceed to model the main isotopic water lines. 
The H$_2$O abundance is kept constant in our models. We have tried models with an abundance increase in the inner region where $T>100$~K, but the current data do not favor these models above the constant-abundance models.

We estimate the absolute uncertainty in the retrieved H$_2$O abundance to be a factor of 10. Since we use the same modeling strategy as the studies by \cite{dr21} and \cite{chavarria}, the \new{abundances obtained} should be comparable to \new{better than a factor of 3}.
Our observed spread in abundances of a factor of $\sim$100 is much larger than this uncertainty.
The same range of abundances is found in other HIFI-based studies of high-mass star-forming regions (\citealt{dr21}; \citealt{chavarria}), and also in previous work with ISO \citep{boonman2003} and from the ground \citep{vandertak2006}.

\new{In conclusion, for the massive star forming regions described in this letter, we clearly detect the contribution of the envelope within the dense core. It is limited to a strong self-absorbed feature, mainly seen in the ground-state line. In order to evaluate it, we first have to remove  emission from outflow shocks and absorption by foreground clouds along the line of sight. The velocities of the absorbers indicate that some are part of the close environment of the source, while others are physically unrelated. 
The derived massive dense core abundances suggest a strong freeze-out of water on dust grains, and imply that water plays only a minor role in the thermal balance of the gas.}

%
The \hho\ line profiles do not seem to depend on the supposed evolutionary stage of the source. For example, the two `hot molecular cores' G31.41 and G29.96 show very different line profiles, and also their \hho\ abundances differ by a factor of $\sim$100. Also, the abundance variations that we have found do not seem related to the luminosity of the sources, their temperature or their turbulent velocity field. However, the number of cases treated is not sufficient for a statistical treatment. Future studies following the same procedure with a larger number of sources should conclude on this issue.
Within our sample, the highest \hho\ abundances are derived for G31.41 and W43-MM1 which show self-absorbed $2_{02}-1_{11}$ line profiles (Fig.~\ref{fig:obs}). As these sources are not the most luminous, hot, or active ones in our sample, the origin of such a high abundance is unclear. 

\new{Firm conclusions about a link between water emission behaviour and the evolutionary stage of the source are limited by the small number
of sources. Our data show that water is a useful tool to understand the gas dynamics in and around massive stars forming regions. Future multi-line studies of larger samples are highly promising to answer key questions about the formation of massive stars.}



\begin{table*}
 \centering
 \caption{Model parameters and derived water abundances.}
 \label{tab:m-pars}
 \begin{center}
  \begin{tabular}{lccccccccc}
  \hline
Source & $M_\mathrm{gas}$ & $r_\mathrm{min}$ & $r_\mathrm{max}$ & $n(r_\mathrm{min})$ & $n(r_\mathrm{max})$ & $T(r_\mathrm{min})$ & $T(r_\mathrm{max})$ & $X_\mathrm{H_2O}$ & $\varv_\mathrm{turb}$\\
     & (\msol) & (AU) & (AU) & (\cmcube) & (\cmcube) & (K) & (K) & & (\kms) \\
  \hline
  \hline
G31.41+0.31 	& 1500	& 200	& 22\,515	& 8.1\ttp{8}	& 3.1\ttp{6}	& 406	& 43.2	& 3.1\ttp{-8} & 1.4 \\
G29.96-0.02 	& 700	& 200	& 20\,700	& 4.4\ttp{8}	& 1.9\ttp{6}	& 489	& 50.8	& $<$5.0\ttp{-10} & 1.1 \\
W33A 		& 4000	& 200	& 62\,000	& 3.5\ttp{8}	& 4.0\ttp{5}	& 291	& 26.0	& 6.0\ttp{-10} & 1.6 \\
W43-MM1 	& 2000	& 200	& 27\,500	& 5.0\ttp{8}	& 2.3\ttp{6}	& 243	& 24.7	& 4.0\ttp{-8} & 3.0 \\
DR21\tablefootmark{a} & 1650	& 2000	& 60\,520	& 1.6\ttp{7}	& 1.5\ttp{5}	& 117	& 23.3	& 2.0\ttp{-10}	& 3.0 \\
W3-IRS5\tablefootmark{b} & 250	& 200	& 12\,000	& 2.9\ttp{8}	& 2.7\ttp{6}	& 480	& 54.7	& 2.0\ttp{-8}	& 2.0 \\
\hline
  \end{tabular}
 \end{center}
\tablefoot{
\tablefoottext{a}{Values from \cite{dr21}}
\tablefoottext{b}{Values from \cite{chavarria}}
}
\end{table*}

\acknowledgement{HIFI has been designed and built by a consortium of institutes and university departments from across Europe, Canada and the United States under the leadership of SRON Netherlands Institute for Space Research, Groningen, The Netherlands and with major contributions from Germany, France and the US. Consortium members are: Canada: CSA, U.Waterloo; France: CESR, LAB, LERMA, IRAM; Germany: KOSMA, MPIfR, MPS; Ireland, NUI Maynooth; Italy: ASI, IFSI-INAF, Osservatorio Astrofisico di Arcetri- INAF; Netherlands: SRON, TUD; Poland: CAMK, CBK; Spain: Observatorio Astron\'{o}mico Nacional (IGN), Centro de Astrobiolog\'{i}a (CSIC-INTA). Sweden: Chalmers University of Technology - MC2, RSS \& GARD; Onsala Space Observatory; Swedish National Space Board, Stockholm University - Stockholm Observatory; Switzerland: ETH Zurich, FHNW; USA:
Caltech, JPL, NHSC.

HIPE is a joint development by the Herschel Science Ground Segment Consortium, consisting of ESA, the NASA Herschel Science Center, and the HIFI, PACS and SPIRE consortia.
}

\bibliography{biblio}
\bibliographystyle{aa}

\newpage

\appendix
\label{appendix}

\section{Massive dense core component extraction}

The velocity profiles of the \hho\ $1_{11}-0_{00}$ line show absorption features  at several velocities. These absorption features arise in foreground clouds along the line of sight or in cold clouds in the neighbourhood of the massive dense core, and are not saturated unlike the absorption from the massive \new{envelope}. In addition to these absorptions, some sources show \hho\ emission from protostellar outflows.

\begin{figure*}[ht]
 \centering
 \resizebox{0.9\hsize}{!}{\includegraphics{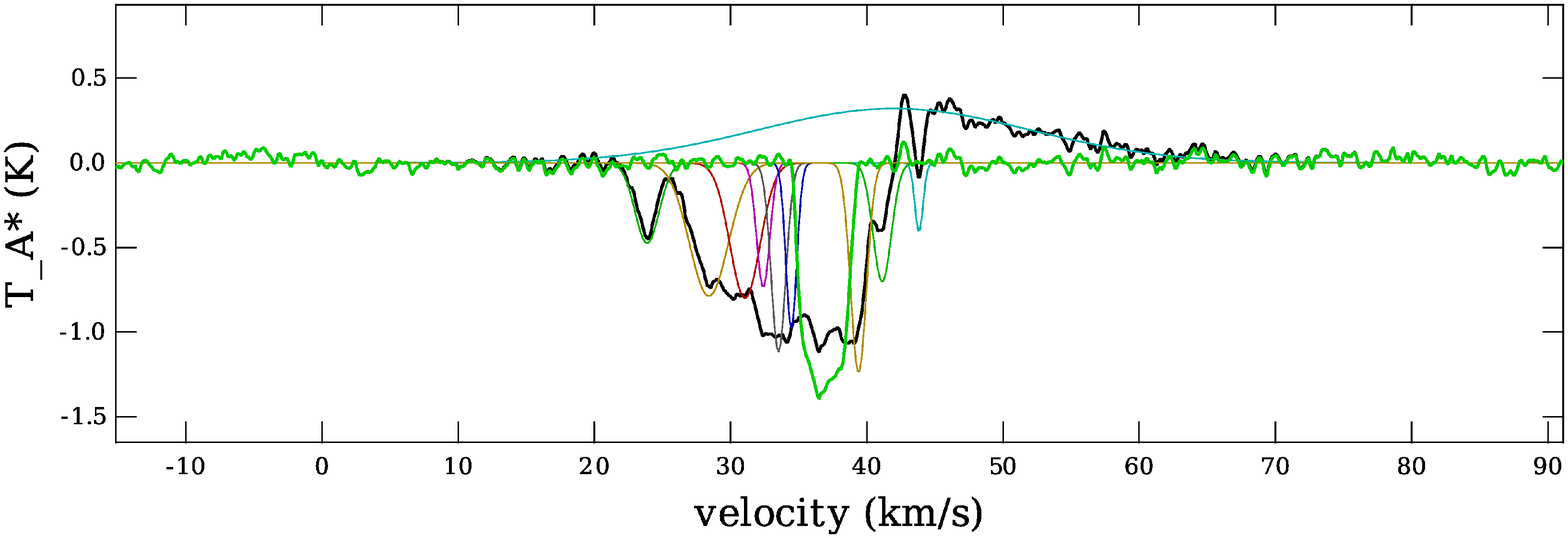}}
 \caption{Extraction of the saturated absorption of para-H$_2$O\,$1_{11}-0_{00}$ line in W33A. Original profile appears in black bold, residual in green bold.} 
 \label{fig:extract-w33a}
\end{figure*}

\begin{figure*}[ht]
 \centering
 \resizebox{0.9\hsize}{!}{\includegraphics{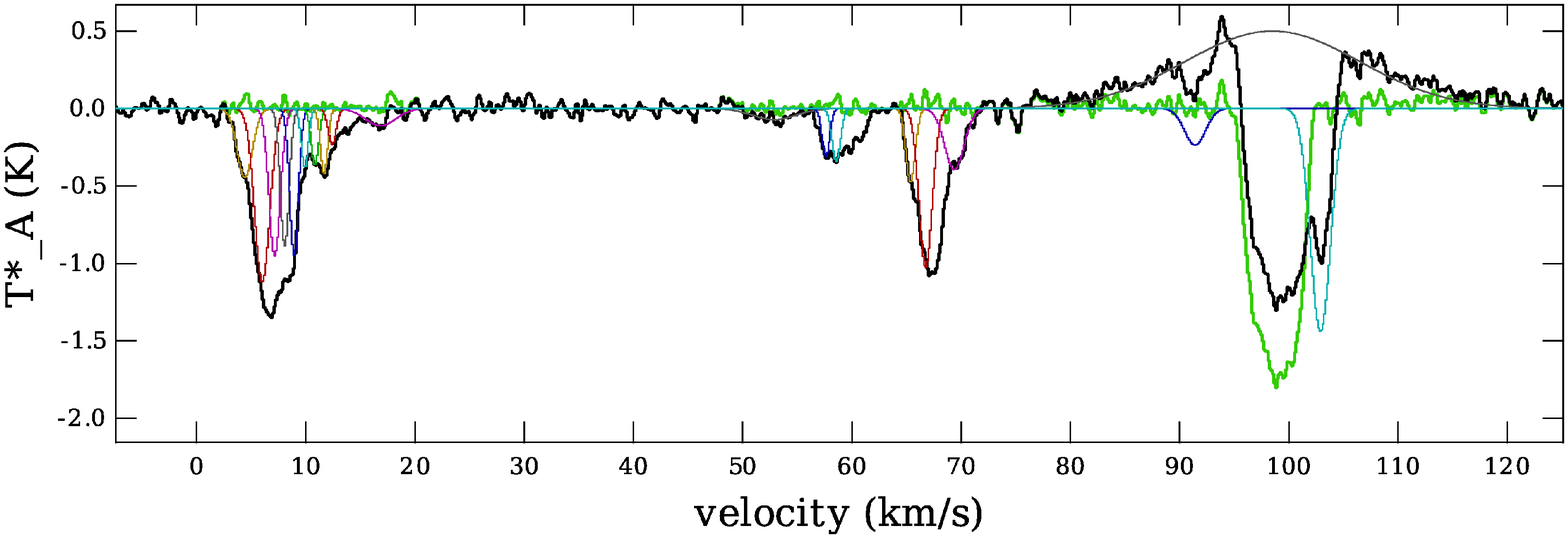}}
 \caption{Extraction of the saturated absorption of para-H$_2$O\,$1_{11}-0_{00}$ line in G29.96. Original profile appears in black bold, residual in green bold.} 
 \label{fig:extract-g29}
\end{figure*}

\begin{figure*}[ht]
 \centering
 \resizebox{0.9\hsize}{!}{\includegraphics{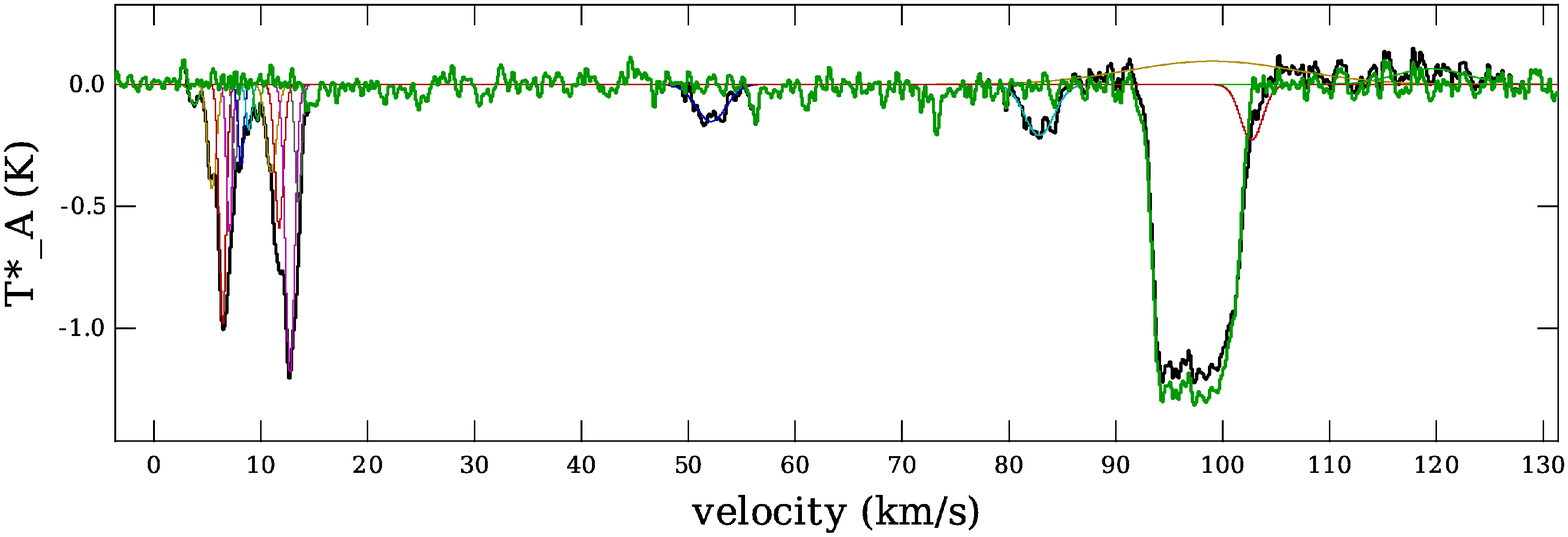}}
 \caption{Extraction of the saturated absorption of para-H$_2$O\,$1_{11}-0_{00}$ line in G31.41. Original profile appears in black bold, residual in green bold.} 
 \label{fig:extract-g31}
\end{figure*}

This appendix presents our procedure to remove these features in order to extract the contribution from the envelope to the line profile. Contrary to others, \new{absorption} from this part of the object is saturated. We are then able to remove other features by iterative Gaussian fits. This process is helped by the high velocity resolution provided by the Herschel-HIFI instrument, showing accurate and "bumpy" profiles in absorptions. Assuming that each bump corresponds to a velocity component, they are removed using the Gaussian fitting tool available in the HIPE software. Starting from the component with the lowest velocity, they are extracted one by one, using the residual of the previous removal to fit the next one. This way of fitting insures a very good extraction of velocity component, giving a quasi-unique final decomposition of the absorption features. Results of this process are given in Fig.~\ref{fig:extract-w43-mm1}, \ref{fig:extract-w33a}, \ref{fig:extract-g29},  \ref{fig:extract-g31} and Tables~\ref{tab:ex-w43-mm1}, \ref{tab:ex-w33a}, \ref{tab:ex-g29} and \ref{tab:ex-g31-41}.


\begin{table}
 \centering
 \caption{Gaussian fit parameters for the full extraction of the saturated absorptions of para-H$_2$O\,$1_{11}-0_{00}$ line in W43-MM1.}
 \label{tab:ex-w43-mm1}
 \begin{center}
  \begin{tabular}{lccc}
  \hline
Component & $T^*_{A}$ & FHWM & $\varv_\mathrm{lsr}$ \\
   \#  & (K) & (\kms) & (\kms) \\
  \hline
  \hline
1 & -0.18 & 1.46 & 62.31 \\
2 & -0.16 & 1.50 & 64.90 \\
3 & -0.20 & 0.66 & 65.57 \\
4 & -0.63 & 1.22 & 66.41 \\
5 & -0.87 & 1.09 & 67.06 \\
6 & -0.36 & 0.71 & 67.75 \\
7 & -0.33 & 1.02 & 68.37 \\
8 & -0.22 & 0.82 & 69.20 \\
9 & -0.53 & 1.19 & 70.09 \\
10 & -0.65 & 1.10 & 70.87 \\
11 & -0.17 & 0.74 & 71.51 \\
12 & -0.09 & 2.46 & 72.25 \\
13 & -0.09 & 1.31 & 74.27 \\
14 & -0.40 & 1.40 & 75.67 \\
15 & -0.54 & 2.35 & 77.29 \\
16 & -0.83 & 1.89 & 78.64 \\
17 & -1.40 & 1.95 & 79.72 \\
18 & -0.80 & 1.59 & 81.04 \\
19 & -0.96 & 1.46 & 82.10 \\
20 & -0.64 & 1.19 & 82.89 \\
21 & -0.32 & 0.71 & 83.51 \\
22 & -0.21 & 2.95 & 87.69 \\
23 & -0.24 & 1.11 & 84.23 \\
24 & -0.13 & 1.52 & 85.33 \\
25 & 0.42 & 23.94 & 99.94 \\
26 & -0.35 & 13.95 & 94.55 \\
27 & -0.58 & 3.78 & 92.48 \\
28 & -1.03 & 2.68 & 94.34 \\
29 & -1.82 & 2.08 & 95.84 \\
30 & -0.38 & 2.03 & 105.42 \\
31 & -1.18 & 1.95 & 104.01 \\
32 & -1.45 & 1.60 & 102.86 \\
33 & -1.68 & 1.55 & 101.83 \\
\hline
\end{tabular}
\end{center}
\end{table}

\begin{table}
 \centering
 \caption{Gaussian fit parameters for for the full extraction of the saturated absorptions of para-H$_2$O\,$1_{11}-0_{00}$ line in W33A.}
 \label{tab:ex-w33a}
 \begin{center}
  \begin{tabular}{lccc}
  \hline
Component & $T^*_{A}$ & FHWM & $\varv_\mathrm{lsr}$ \\
   \#  & (K) & (\kms) & (\kms) \\
  \hline
  \hline
1 & 0.32 & 32.95 & 42.00 \\
2 & -0.48 & 2.92 & 23.87 \\
3 & -0.79 & 4.70 & 28.41 \\
4 & -0.80 & 3.64 & 31.06 \\
5 & -0.74 & 1.60 & 32.40 \\
6 & -1.12 & 1.75 & 33.52 \\
7 & -0.98 & 1.31 & 34.46 \\
8 & -0.41 & 1.00 & 43.83 \\
9 & -0.70 & 2.20 & 41.13 \\
10 & -1.24 & 1.91 & 39.40 \\
\hline
\end{tabular}
\end{center}
\end{table}

\begin{table}
 \centering
 \caption{Gaussian fit parameters for for the full extraction of the saturated absorptions of para-H$_2$O\,$1_{11}-0_{00}$ line in G29.96.}
 \label{tab:ex-g29}
 \begin{center}
  \begin{tabular}{lccc}
  \hline
Component & $T^*_{A}$ & FHWM & $\varv_\mathrm{lsr}$ \\
   \#  & (K) & (\kms) & (\kms) \\
  \hline
  \hline
1 & -0.44 & 2.32 & 4.41 \\
2 & -1.12 & 2.30 & 6.01 \\
3 & -0.95 & 1.71 & 7.15 \\
4 & -0.89 & 1.30 & 8.08 \\
5 & -0.95 & 1.32 & 8.93 \\
6 & -0.38 & 1.04 & 9.91 \\
7 & -0.37 & 1.14 & 10.84 \\
8 & -0.43 & 1.06 & 11.68 \\
9 & -0.23 & 1.21 & 12.42 \\
10 & -0.11 & 4.55 & 16.85 \\
11 & -0.07 & 6.52 & 52.89 \\
12 & -0.31 & 1.13 & 57.62 \\
13 & -0.34 & 1.29 & 58.57 \\
14 & -0.25 & 3.12 & 60.10 \\
15 & -0.48 & 1.39 & 65.37 \\
16 & -1.03 & 2.00 & 66.76 \\
17 & -0.40 & 2.93 & 69.38 \\
18 & 0.50 & 26.36 & 98.50 \\
19 & -0.24 & 3.17 & 91.42 \\
20 & -1.44 & 3.06 & 102.91 \\
\hline
\end{tabular}
\end{center}
\end{table}

\begin{table}
 \centering
 \caption{Gaussian fit parameters for for the full extraction of the saturated absorptions of para-H$_2$O\,$1_{11}-0_{00}$ line in G31.41.}
 \label{tab:ex-g31-41}
 \begin{center}
  \begin{tabular}{lccc}
  \hline
Component & $T^*_{A}$ & FHWM & $\varv_\mathrm{lsr}$ \\
   \#  & (K) & (\kms) & (\kms) \\
  \hline
  \hline
1 & -0.08 & 1.36 & 3.79 \\
2 & -0.43 & 1.35 & 5.44 \\
3 & -0.99 & 1.06 & 6.44 \\
4 & -0.61 & 0.86 & 7.06 \\
5 & -0.35 & 0.68 & 7.57 \\
6 & -0.34 & 0.84 & 8.15 \\
7 & -0.19 & 0.88 & 8.89 \\
8 & -0.36 & 1.61 & 10.99 \\
9 & -0.59 & 1.31 & 11.72 \\
10 & -1.19 & 1.40 & 12.75 \\
11 & -0.49 & 0.98 & 13.53 \\
12 & -0.15 & 4.63 & 52.14 \\
13 & -0.21 & 4.53 & 82.84 \\
14 & 0.06 & 10.92 & 119.53 \\
15 & 0.10 & 25.85 & 99.02 \\
16 & -0.23 & 3.41 & 102.70 \\
\hline
\end{tabular}
\end{center}
\end{table}

\end{document}